\documentclass[11pt]{article}
\usepackage{xeCJK}
\usepackage{xcolor}
\usepackage{mathtools}
\usepackage{amsmath}
\usepackage{amssymb}
\usepackage{graphicx}
\usepackage{geometry}
\usepackage[bookmarks=false,
 breaklinks=false,pdfborder={0 0 1},backref=false,colorlinks=true]
 {hyperref}
\hypersetup{
 linkcolor=black,citecolor=black,urlcolor=black}

\makeatletter
\@ifundefined{date}{}{\date{}}

\usepackage{bm}

\usepackage{booktabs}

\usepackage{xcolor}

\usepackage{tikz}
\usetikzlibrary{arrows.meta,positioning}

\usepackage[nospace]{cite}

\usepackage{etoolbox}
\AtBeginEnvironment{thebibliography}{
  \small\raggedright\sloppy
  \setlength{\itemsep}{0pt plus 0.1ex}
  \setlength{\parsep}{0pt}
}

\usepackage[nameinlink]{cleveref}

\usepackage{caption}
\usepackage{placeins}
\usepackage{float}

\title{Finite-volume Analysis of On-shell Subtraction in the Lüscher formalism\thanks{This work was supported by the National Natural Science Foundation of China (NSFC) under Grant No. 12135011.}}

\author{Wanchen Lei, Yuting Wang, Jin-Yi Pang\thanks{Corresponding author. Email: jypang@usst.edu.cn}\\\\{\em School of Physics,  Faculty of Basic Sciences}\\\\{\em University of Shanghai for Science and Technology, Shanghai 200093, China}}

\makeatother

\begin{document}
\maketitle 
\begin{abstract}
\noindent On-shell subtraction is an important step in finite-volume analyses, in which an on-shell singularity is isolated and regular momentum-dependent factors are evaluated at their on-shell values. For short-range interactions with a finite mass gap, the correction associated with this on-shell subtraction is exponentially suppressed at sufficiently large volumes and can therefore be neglected at the corresponding level of accuracy. We examine how this correction changes as the relevant mass parameter decreases. Using a standard dimensional-regularization prescription, we derive analytic expressions for the resulting finite-volume corrections for representative momentum structures and investigate their behavior numerically. For a nonzero mass parameter, the corrections are indeed exponentially suppressed at large $\mu L$, whereas, in the massless limit, this suppression is replaced by an enhanced finite-volume dependence. More importantly, the $S$-wave contribution can develop an infrared divergence in the finite-volume correction in the massless limit. These results clarify the conditions under which on-shell subtraction can be applied and provide guidance for assessing such corrections in finite-volume hadronic calculations involving long-range effects, including QCD+QED.
\end{abstract}

\section{Introduction}

Finite-volume methods exploit the volume dependence of hadronic correlation
functions and spectra to constrain infinite-volume amplitudes \cite{ref48,ref49}. In
the two-particle sector, Lüscher's formalism relates the discrete
spectrum in a cubic box to the infinite-volume scattering amplitude
or, in the appropriate kinematic regime, to elastic scattering phase
shifts \cite{ref31,ref13,ref25}. Moving-frame extensions and related
quantization conditions establish the corresponding relations for moving
frames and more general kinematics \cite{ref14,ref15,ref26,ref27,ref28}.
Generalizations to coupled channels, arbitrary spin, and matrix elements
retain closely related finite-volume kinematic functions and pole
structures \cite{ref32,ref33,ref34,ref35,ref36}. Despite their distinct
physical settings, these constructions share a two-particle on-shell
singularity whose sum$-$integral difference produces the leading power-law
volume dependence.

The volume dependence is controlled by the analytic structure of the
summand. If the pole-subtracted summand is analytic in the relevant
momentum domain and is separated from the nearest singularity by a
nonzero mass scale, its sum$-$integral difference is exponentially
suppressed at large volume, with the exponent determined by the nearest
relevant singularity \cite{ref1,ref2,ref3,ref4,ref5}. By contrast,
the two-particle on-shell pole generates the power-law finite-volume
dependence in the Lüscher formula \cite{ref31,ref13}. After dimensionless
rescaling, the singular part is governed by the denominator $(n^{2}-\eta^{2})^{-1}$,
where $\mathbf{n}=\mathbf{k}L/(2\pi)$, $n=|\mathbf{n}|$, and
$\eta=q_{\mathrm{on}}L/(2\pi)$. Here, $L$ is the side length of the
periodic cubic box, $\mathbf{k}$ is the relative momentum, and
$q_{\mathrm{on}}$ is its on-shell magnitude. The principal-value
sum$-$integral difference associated with this pole enters the finite-volume
loop function, or kinematic $F$ function, that appears in the quantization
condition \cite{ref13,ref25,ref14,ref15,ref34}. A smooth momentum-dependent
factor multiplying this pole is normally evaluated at the on-shell
momentum \cite{ref46,ref47}. This procedure preserves the leading power-law term. A recent
moving-frame derivation makes this separation explicit and demonstrates
that, under the stated analyticity assumptions, the sum$-$integral
difference of the pole-subtracted part is exponentially suppressed.
Within the same framework, three boost prescriptions yield the same
power-law behavior of the finite-volume energy spectrum, with their
differences restricted to exponentially suppressed terms \cite{ref45}.

Recent studies have extended finite-volume scattering formalisms to
systems in which light-particle exchange generates long-range interactions
and nearby left-hand singularities \cite{ref41,ref42,ref43,ref44}.
These approaches modify the finite-volume quantization condition to
treat the exchange dynamics and associated singularities explicitly.
The question addressed here is complementary: we isolate the sum$-$integral
correction generated when momentum-dependent factors that are regular
at the two-particle pole are evaluated at their on-shell values and
determine how this correction changes as the relevant mass scale
approaches the massless limit.

This work determines this sum$-$integral difference. A direct example
is provided by the relativistic two-particle propagator: after the
energy integration, its smooth momentum dependence contains
$1/\omega_{\mathbf{k}}$, where $\omega_{\mathbf{k}}=\sqrt{k^{2}+m_{h}^{2}}$
is the single-hadron energy, $k=|\mathbf{k}|$, and $m_{h}$ is the hadron
mass. In dimensionless variables,
$\omega_{\mathbf{k}}=(2\pi/L)E_{n}$, with
$E_{n}=\sqrt{n^{2}+\nu^{2}}$ and $\nu=m_{h}L/(2\pi)$.
Thus, apart from the overall factor $L/(2\pi)$, the inverse-energy
factor becomes $1/E_{n}$. More generally, $g(n;\nu)$ denotes a radial
momentum-dependent factor that is regular at $n=\eta$, and the on-shell
step replaces it with its value $g(\eta;\nu)$ in the pole term.
We examine how the associated sum$-$integral difference depends on
the radial momentum structure, the dimensionless mass parameter $\nu$,
the on-shell parameter $\eta$, and the partial-wave component.

We analyze four momentum dependences: $n^{2s}$, $n^{2s+1}$, $n^{2s}E_{n}$,
and $n^{2s}/E_{n}$. We first consider the even- and odd-power cases.
The cases involving $E_{n}$ and $1/E_{n}$ retain the relativistic
factors that arise in finite-volume loop calculations. A partial-wave
expansion, a common Cauchy principal-value prescription, and dimensional
regularization place all four cases within a unified framework. For the
relativistic two-hadron loop motivating this analysis, $\nu=m_{h}L/(2\pi)>0$,
where $m_{h}$\ is the hadron mass. In the general analysis below, we
write $\nu=\mu L/(2\pi)$, where $\mu$ denotes the mass parameter
entering the momentum-dependent structure under consideration. The
two-hadron propagator therefore corresponds to $\mu=m_{h}$. We also
examine the formal limit $\nu\to0^{+}$, which describes the massless
limit of the corresponding momentum structure. The specific contribution
of this work is the explicit evaluation of the on-shell-subtraction
corrections for these four radial structures. This evaluation includes
a unified decomposition, the connection between the massive $E_{n}$
and $1/E_{n}$ structures and their odd-power limits where those limits
are finite, and the exceptional $s=0$, $l=0$ behavior of the $1/E_{n}$
case when the zero-momentum mode is retained. For the even-power polynomial
class, the vanishing of the on-shell-subtraction sum$-$integral term
is specific to the dimensional-regularization finite-part prescription
adopted in this work.

The limit $\nu\to0^{+}$ at fixed $L$ becomes relevant when the
corresponding momentum dependence is generated by a genuinely massless
propagating degree of freedom. Finite-volume QCD+QED is one possible
application because the photon produces long-range finite-volume effects,
and the infrared sector depends on the chosen formulation \cite{ref6,ref7,ref8,ref9,ref40}.
In this setting, the sum$-$integral differences derived here enter
the finite-volume calculation together with the infrared prescription
and the process-dependent charge, spin, gauge, and matching structures.

The paper is organized as follows. Section 2 derives the on-shell
subtraction from a relativistic two-particle loop and defines the
sum$-$integral difference studied below. Section 3 develops the analytic
results for the four momentum dependences and examines their behavior
as $\nu\to0^{+}$. Section 4 presents the numerical dependence on
mass, on-shell momentum, and partial-wave component. Section 5 discusses
the physical interpretation of these functions and their incorporation
into specified finite-volume calculations, and Sec.~6 summarizes
the main results.

\section{On-shell subtraction and the sum$-$integral difference in the Lüscher
formalism}

We briefly review the steps in the derivation of the Lüscher formula
that are relevant to the on-shell subtraction. Consider the elastic
scattering of two identical spinless hadrons in a periodic cubic volume
$L^{3}$ at vanishing total momentum and below the first inelastic threshold.
The finite-volume scattering equation is obtained by replacing the
loop-momentum integral with a discrete sum,

\begin{align}
T^{(L)}(\mathbf{p},\mathbf{q};q^{2}_{\mathrm{on}}) & =V(\mathbf{p},\mathbf{q})\nonumber \\
 & +\frac{1}{L^{3}}\sum_{\mathbf{k}}V(\mathbf{p},\mathbf{k})G(k;q^{2}_{\mathrm{on}})T^{(L)}(\mathbf{k},\mathbf{q};q^{2}_{\mathrm{on}}).\label{eq:finite-volume-scattering-equation}
\end{align}
Here, $\mathbf{k}=2\pi\mathbf{n}/L$ denotes the discrete loop momentum, whereas the external momenta are treated as continuous variables. The poles of $T^{(L)}$ determine the finite-volume energy spectrum \cite{ref31,ref13,ref25}. Before proceeding to the finite-volume analysis, we write the corresponding infinite-volume equation as

\begin{align}
T(p,q;q^{2}_{\mathrm{on}}) & =V(p,q)\nonumber \\
 & +\int\frac{d^{3}k}{(2\pi)^{3}}\,V(p,k)G(k;q^{2}_{\mathrm{on}})T(k,q;q^{2}_{\mathrm{on}}).
\end{align}
Above threshold, the real part of the two-particle pole contribution
is understood in the principal-value sense. In a finite volume, the
loop integral is replaced by the momentum sum in Eq.~\eqref{eq:finite-volume-scattering-equation}.
Its finite-volume dependence can be isolated through

\begin{equation}
\begin{aligned}\frac{1}{L^{3}}\sum_{\mathbf{k}}= & \left[\frac{1}{L^{3}}\sum_{\mathbf{k}}-\text{PV}\int\frac{d^{3}\mathbf{k}}{(2\pi)^{3}}\right]\\
 & +\text{PV}\int\frac{d^{3}\mathbf{k}}{(2\pi)^{3}}.
\end{aligned}
\label{eq:sum-integral-separation}
\end{equation}
After isolating the two-particle pole, the regular momentum dependence
can be separated from its value at $k=q_{\mathrm{on}}$. In the partial-wave
representation, the corresponding short-range relation is given by
\begin{align}
 & \frac{1}{L^{3}}\sum_{\mathbf{k}}\left(\frac{\mathcal{Y}^{*}_{\ell m}(\mathbf{k})\mathcal{Y}_{jn}(\mathbf{k})}{k^{2}-q^{2}_{\mathrm{on}}}\right)\left[V_{\ell}(p,k)R^{(L)}_{jn,\ell'm'}(k,q)-V_{\ell}(p,q_{\mathrm{on}})R^{(L)}_{jn,\ell'm'}(q_{\mathrm{on}},q)\right]\nonumber \\
= & \int\dfrac{d^{3}\mathbf{k}}{(2\pi)^{3}}\left(\frac{\mathcal{Y}^{*}_{\ell m}(\mathbf{k})\mathcal{Y}_{jn}(\mathbf{k})}{k^{2}-q^{2}_{\mathrm{on}}}\right)\left[V_{\ell}(p,k)R^{(L)}_{jn,\ell'm'}(k,q)-V_{\ell}(p,q_{\mathrm{on}})R^{(L)}_{jn,\ell'm'}(q_{\mathrm{on}},q)\right]+O(e^{-L/r}).\label{eq:regular-sum-to-integral}
\end{align}
For a finite-range interaction of range $r$ and a pole-subtracted integrand
that is analytic in the loop momentum, Eq.~\eqref{eq:regular-sum-to-integral}
provides the short-range input used in the standard Lüscher derivation
\cite{ref31,ref13,ref25}. It states that the sum$-$integral difference
of the pole-subtracted contribution is exponentially suppressed, such that the
power-law finite-volume dependence arises from the isolated on-shell
term. This conclusion does not necessarily hold when massless propagation
alters the infrared behavior of the pole-subtracted expression \cite{ref40}.

We now identify the specific factor examined in this work. For two
identical relativistic particles of mass $m_{h}$, define

\begin{equation}
\omega_{\mathbf{k}}=\sqrt{k^{2}+m^{2}_{h}},\qquad E=2\omega_{q_{\mathrm{on}}}.\label{eq:relativistic-energy-definitions}
\end{equation}
With the definitions in Eq.~\eqref{eq:relativistic-energy-definitions},
the relativistic scattering equation, after integration over the loop energy,
contains

\begin{equation}
G(k;q^{2}_{\mathrm{on}})=\frac{E+2\omega_{\mathbf{k}}}{8\omega_{\mathbf{k}}E\left(k^{2}-q^{2}_{\mathrm{on}}\right)}.
\end{equation}
The loop factors under the normalization conventions are related by

\begin{equation}
\frac{E+2\omega_{\mathbf{k}}}{8\omega_{\mathbf{k}}E\left(k^{2}-q^{2}_{\mathrm{on}}\right)}=-\frac{1}{E}\,\frac{1}{2\omega_{\mathbf{k}}\left(E-2\omega_{\mathbf{k}}\right)}.\label{eq:relativistic-propagator-conventions}
\end{equation}
The factor studied here therefore follows directly from the relativistic
two-particle loop in Lüscher's derivation. The two-particle pole
is isolated before the regular momentum-dependent factors multiplying
it are evaluated at the on-shell momentum. Equation~\eqref{eq:relativistic-propagator-conventions}
shows that the two loop normalizations differ only by a $k$-independent
factor. They therefore have the same loop-momentum dependence
relevant to the on-shell subtraction.

The present analysis isolates the sum$-$integral difference arising
from this loop-momentum dependence. The interaction-dependent factors
are not included in the functions studied below; their role in complete
finite-volume hadronic calculations is discussed in Sec. 5.

For the polynomial and relativistic momentum structures considered
below, both the finite-volume sum and the corresponding infinite-volume
integral may be ultraviolet divergent. Their finite parts are
therefore defined using the same dimensional-regularization prescription,
ensuring that the sum$-$integral difference is evaluated consistently \cite{ref13,ref25,ref14,ref15,ref34}.
To explicitly display the loop-momentum dependence before the on-shell subtraction,
the second loop normalization in Eq.~\eqref{eq:relativistic-propagator-conventions}
can be rewritten as 

\begin{equation}
\frac{1}{2\omega_{\mathbf{k}}\left(E-2\omega_{\mathbf{k}}\right)}=\frac{\omega_{q_{\mathrm{on}}}+\omega_{\mathbf{k}}}{4\omega_{\mathbf{k}}\left(q^{2}_{\mathrm{on}}-k^{2}\right)}.\label{eq:two-particle-propagator-pole}
\end{equation}

Equation~\eqref{eq:two-particle-propagator-pole} makes the two-particle
on-shell pole explicit through the factor $(q^{2}_{\mathrm{on}}-k^{2})^{-1}$.
The prefactor multiplying the pole, $(\omega_{q_{{\rm on}}}+\omega_{\mathbf{k}})/(4\omega_{\mathbf{k}})$,
is regular at $|\mathbf{k}|=q_{\mathrm{on}}$ and depends nontrivially
on the loop momentum only through $(\omega_{\mathbf{k}})^{-1}$. To
express this factor on the dimensionless momentum lattice $\mathbf{n}\in\mathbb{Z}^{3}$,
we introduce
\begin{equation}
\eta=\frac{q_{\mathrm{on}}L}{2\pi},\qquad\nu=\frac{m_{h}L}{2\pi}.\label{eq:dimensionless-parameters}
\end{equation}
With the parameters defined in Eq.~\eqref{eq:dimensionless-parameters}, define

\begin{equation}
E_{n}=\sqrt{n^{2}+\nu^{2}},\qquad E_{\eta}=\sqrt{\eta^{2}+\nu^{2}}.\label{eq:dimensionless-energies}
\end{equation}
For the specific two-hadron loop considered in this section, the general
mass parameter defined below is therefore identified as $\mu=m_{h}$.
Using the energies in Eq.~\eqref{eq:dimensionless-energies}, the
prefactor multiplying the pole becomes

\begin{equation}
\frac{\omega_{q_{{\rm on}}}+\omega_{\mathbf{k}}}{4\omega_{\mathbf{k}}}=\frac{E_{n}+E_{\eta}}{4E_{n}}.
\end{equation}
Subtracting the on-shell value of \(1/2\) from this prefactor gives

\begin{equation}
\frac{E_{n}+E_{\eta}}{4E_{n}}-\frac{1}{2}=\frac{E_{\eta}}{4}\left(\frac{1}{E_{n}}-\frac{1}{E_{\eta}}\right).\label{eq:dimensionless-on-shell-difference}
\end{equation}
Equation~\eqref{eq:dimensionless-on-shell-difference} shows that the loop-momentum dependence discarded by the on-shell subtraction is proportional to $1/E_{n}-1/E_{\eta}$. Section~3 therefore considers the term containing $1/E_{n}$ together with the other momentum structures required to analyze the on-shell subtraction.

\section{Finite-volume corrections to the on-shell subtraction}

 The on-shell subtraction can be formulated in terms of a function $g(n;\nu)$, with $\nu=\mu L/(2\pi)$, that multiplies the standard two-particle pole in the finite-volume loop \cite{ref31,ref13,ref25,ref15,ref34}. This subtraction amounts to replacing $g(n;\nu)$ with its on-shell value~$g(\eta;\nu)$. Using the partial-wave expansion, the corresponding sum$-$integral difference can be written as

\begin{equation}
\mathcal{D}_{lm}(\eta,\nu)=\left(\sum_{\mathbf{n}\in\mathbb{Z}^{3}}-\text{PV}\int_{\mathbb{R}^{3}}d^{3}\mathbf{n}\right)\frac{\left[g(n;\nu)-g(\eta;\nu)\right]\mathcal{Y}_{lm}}{n^{2}-\eta^{2}},\label{eq:Dg-general}
\end{equation}
where
\begin{equation}
\mathcal{Y}_{lm}(\mathbf{n})=n^{l}Y_{lm}(\hat{\mathbf{n}}),\qquad n=|\mathbf{n}|.
\end{equation}
In the traditional Lüscher treatment of short-range hadronic interactions,
the power-law finite-volume dependence arises from the on-shell pole of the
two-hadron propagator. Momentum-dependent terms that are regular at this pole
may therefore be evaluated on shell, since the resulting difference contributes
only exponentially suppressed finite-volume effects. Here, we retain this
difference and examine it for the following four choices:

\begin{equation}
g(n;\nu)=n^{2s},\qquad n^{2s+1},\qquad n^{2s}E_{n},\qquad\frac{n^{2s}}{E_{n}},\qquad s\in\mathbb{N}_{0}.
\end{equation}

These four choices are used to examine how different radial momentum
dependences affect the finite-volume correction associated with the
on-shell subtraction. The even-power form $n^{2s}$ provides an analytic
benchmark for short-range behavior \cite{ref25}, whereas the odd-power form $n^{2s+1}$,
which is nonanalytic in $n^{2}$, serves as a prototype for algebraically
decaying finite-volume corrections associated with long-range effects
\cite{ref40}. The form $n^{2s}E_{n}$ introduces a finite mass through the
relativistic dispersion relation and reduces to $n^{2s+1}$ in the massless
limit $\nu\to0$. Finally, $n^{2s}/E_{n}$ incorporates the $1/E_{n}$ dependence
associated with the relativistic two-particle propagator discussed in
Sec.~2 \cite{ref15}, with $n^{2s}$ representing additional analytic momentum dependence.

For these momentum structures, $s$ characterizes the radial momentum dependence,
and the corresponding quantities are denoted by $\mathcal{D}_{s,l}$.
Because $g(n;\nu)$ is radial, the corresponding infinite-volume angular
integral vanishes for $l>0$. In a finite cubic volume with vanishing total
momentum, however, the discrete momentum sum is governed by cubic symmetry
$O_{h}$ rather than continuous rotational symmetry. For sums over complete
cubic shells, all odd-$l$ components and the $l=2$ component vanish,
whereas certain higher even-$l$ components may remain nonzero \cite{ref26}.
In particular, $l=4$ is the first higher angular component permitted by
cubic symmetry and is included in the numerical analysis of Sec.~4.
We therefore retain the general $\mathcal{Y}_{lm}$-projected expressions
to account for higher partial waves permitted by cubic symmetry, although
only the $l=0$ component is required for a pure $S$-wave truncation.
In the moving frame, all partial-wave contributions should be included.
For notational simplicity, the magnetic quantum number $m$ is not shown
explicitly in $\mathcal{D}_{s,l}$; its dependence is retained through
$\mathcal{Y}_{lm}(\mathbf{n})$, and the relations among the permitted magnetic
components are summarized in Appendix~\ref{app:cubic-partial-waves}.

Throughout the following analysis, the finite-volume sums and the
corresponding infinite-volume integrals are defined using the same
dimensional-regularization prescription. For ultraviolet-divergent
expressions, the finite part is extracted after forming the sum$-$integral
difference. The Cauchy principal-value prescription is applied consistently
with the treatment of the original two-particle pole \cite{ref15}.

\subsection{Even-power case }

The even-power case provides the simplest nontrivial test of the substitution
$n^{2s}\to\eta^{2s}$. For $g(n;\nu)=n^{2s}$, the corresponding sum$-$integral
difference is

\begin{equation}
\mathcal{D}^{\mathrm{even}}_{s,l}(\eta)=\left(\sum_{\mathbf{n}}-\mathrm{PV}\int d^{3}\mathbf{n}\right)\frac{(n^{2s}-\eta^{2s})\mathcal{Y}_{lm}(\mathbf{n})}{n^{2}-\eta^{2}}.
\end{equation}
The pole is removed algebraically before taking the finite part:

\begin{equation}
\frac{n^{2s}-\eta^{2s}}{n^{2}-\eta^{2}}=\sum^{s-1}_{j=0}\eta^{2(s-1-j)}n^{2j}.
\end{equation}

Thus, this term is a finite linear combination of polynomial coefficients.
Under the dimensional-regularization finite-part prescription used here,
these complete polynomial moments have a vanishing sum$-$integral finite
part after the cubic-shell projection.

\begin{equation}
\begin{aligned}\mathcal{D}^{\mathrm{even}}_{s,l}(\eta)={} & \sum_{\mathbf{n}}\mathcal{Y}_{lm}(\mathbf{n})e^{\eta^{2}}\Bigg[\frac{(n^{2s}-\eta^{2s})e^{-n^{2}}}{n^{2}-\eta^{2}}\\
 & \qquad{}-\pi^{l+\frac{3}{2}}e^{-\pi^{2}n^{2}}\sum^{s-1}_{j=0}\eta^{2(s-1-j)}j!(-i)^{l}L^{(l+\frac{1}{2})}_{j}(\pi^{2}n^{2})\Bigg],
\end{aligned}
\end{equation}
where the generalized Laguerre polynomial is defined as

\begin{equation}
L^{(l+1/2)}_{j}(\pi^{2}n^{2})=\sum^{j}_{i=0}(-1)^{i}\frac{\Gamma(j+l+3/2)}{i!(j-i)!\Gamma(i+l+3/2)}(\pi^{2}n^{2})^{i}.
\end{equation}
The even-power case describes the regular momentum dependence characteristic of a short-range interaction. Since this contribution is analytic in \(n^{2}\), it can be expanded in even powers of the relative momentum. Under the standard dimensional-regularization finite-part prescription adopted here, it therefore follows that

\begin{equation}
\mathcal{D}^{\mathrm{even}}_{s,l}(\eta)=0.
\end{equation}
This result shows that, for this short-range form, replacing the internal
momentum dependence $n^{2s}$ with its on-shell value~$\eta^{2s}$
does not alter the sum$-$integral difference. Thus, for this polynomial
class, within the adopted dimensional-regularization finite-part prescription, the on-shell
subtraction generates no additional sum$-$integral contribution.

\subsection{Odd-power case}

To analyze the novel, nontrivial structure in the odd-power case, we
first consider the case without the energy factor arising from a relativistic
dispersion relation. In contrast to the complete cancellation observed in the
even-power case, the corresponding sum$-$integral difference does
not vanish for odd-power momentum dependence. We take $s\ge0$ to
be an integer. When the summand has a singularity on the real axis
at $n^{2}=\eta^{2}$, the continuous integral is defined in the Cauchy
principal-value sense. For $g(n;\nu)=n^{2s+1}$, we obtain

\begin{equation}
\mathcal{D}^{\mathrm{odd}}_{s,l}(\eta)=\left(\sum_{\mathbf{n}\in\mathbb{Z}^{3}}-\mathrm{PV}\int_{\mathbb{R}^{3}}d^{3}\mathbf{n}\right)\frac{n^{2s+1}\mathcal{Y}_{lm}(\mathbf{n})-\eta^{2s+1}\mathcal{Y}_{lm}(\mathbf{n})}{n^{2}-\eta^{2}}.
\end{equation}
Decomposing the difference in the numerator, we use the identity

\begin{equation}
\frac{n^{2s+1}-\eta^{2s+1}}{n^{2}-\eta^{2}}=\sum^{s-1}_{j=0}\eta^{2(s-1-j)}n^{2j+1}+\eta^{2s}\frac{1}{n+\eta},
\end{equation}
Substituting this identity into the definition of $\mathcal{D}^{\mathrm{odd}}_{s,l}$ and applying the linearity of the sum-minus-integral operator yields

\begin{equation}
\mathcal{D}^{\mathrm{odd}}_{s,l}(\eta)=\sum^{s-1}_{j=0}\eta^{2(s-1-j)}c^{\mathrm{odd}}_{j,l}+\eta^{2s}\mathcal{R}^{\mathrm{odd}}_{l}(\eta).
\end{equation}
where

\begin{align}
c^{\mathrm{odd}}_{j,l} & =\left(\sum_{\mathbf{n}}-\int d^{3}\mathbf{n}\right)n^{2j+1}\mathcal{Y}_{lm}(\mathbf{n}),\\
\mathcal{R}^{\mathrm{odd}}_{l}(\eta) & =\left(\sum_{\mathbf{n}}-\int d^{3}\mathbf{n}\right)\frac{\mathcal{Y}_{lm}(\mathbf{n})}{n+\eta}.
\end{align}

The coefficients $c^{\mathrm{odd}}_{j,l}$ are independent of the
external momentum parameter $\eta$, and the first term in the decomposition
is a finite polynomial in $\eta^{2}$. The function~$\mathcal{R}^{\mathrm{odd}}_{l}(\eta)$
is independent of $s$ and contains the nonpolynomial dependence
on $\eta$. These two contributions arise from the same algebraic decomposition
of the finite-volume correction associated with the on-shell subtraction.
Because $n^{2s+1}$ is not a polynomial in $n^{2}$, the subtraction does
not reduce the full expression to a finite sum of polynomial moments.
Odd powers of the loop momentum may appear in prefactors with nonanalytic
momentum dependence, including structures associated with long-range
interactions. In such cases, the corresponding finite-volume correction
need not be exponentially suppressed and may instead exhibit algebraic
volume dependence. This behavior falls outside the short-range
assumptions of the traditional Lüscher formalism.

The quantities defined above are not ordinary, absolutely convergent
sums and therefore require regularization. Their finite parts are
defined using dimensional regularization, with the same prescription
applied to both the finite-volume sum and the corresponding infinite-volume
integral. The factor $e^{-\epsilon n^{2}}$ introduced below is used
solely to regulate the intermediate sums and integrals. It is inserted
into both terms and removed by taking the limit $\epsilon\to0^{+}$
after forming the sum$-$integral difference. The odd-power coefficients
can thus be written as
\begin{equation}
c^{\mathrm{odd}}_{j,l}=\lim\limits_{\epsilon\to0^{+}}\left[\sum_{\mathbf{n}\in\mathbb{Z}^{3}}n^{2j+1}\mathcal{Y}_{lm}(\mathbf{n})e^{-\epsilon n^{2}}-\int_{\mathbb{R}^{3}}d^{3}\mathbf{n}\,n^{2j+1}\mathcal{Y}_{lm}(\mathbf{n})e^{-\epsilon n^{2}}\right].
\end{equation}
Similarly, the nonpolynomial term can be represented as
\begin{equation}
\mathcal{R}^{\mathrm{odd}}_{l}(\eta)=\lim\limits_{\epsilon\to0^{+}}\left[\sum_{\mathbf{n}\in\mathbb{Z}^{3}}\frac{\mathcal{Y}_{lm}(\mathbf{n})}{n+\eta}e^{-\epsilon n^{2}}-\int_{\mathbb{R}^{3}}d^{3}\mathbf{n}\,\frac{\mathcal{Y}_{lm}(\mathbf{n})}{n+\eta}e^{-\epsilon n^{2}}\right].
\end{equation}

To derive explicit representations of $c_{j,l}$ and $\mathcal{R}(\eta)$,
we apply the Poisson summation formula to the corresponding regulated
sum--integral differences. In the partial-wave expansion, the angular
integration selects the corresponding $l$th partial wave and yields the
phase factor $(-i)^{l}$, as in standard finite-volume partial-wave
analyses \cite{ref25,ref26,ref27,ref28,ref35}. Applying this procedure
to the regulated odd-power terms and extracting the finite part yields 
\begin{equation}
c^{\mathrm{odd}}_{j,l}=(-i)^{l}\pi^{-\left(2j+l+\frac{5}{2}\right)}\frac{\Gamma(j+l+2)}{\Gamma\!\left(-j-\frac{1}{2}\right)}\sum_{\mathbf{n}\ne0}\frac{\mathcal{Y}_{lm}(\mathbf{n})}{n^{2j+2l+4}}.\label{eq:codd-poisson}
\end{equation}
Equation~\eqref{eq:codd-poisson} shows that, after applying Poisson
summation, the coefficients defined above can be expressed as power-law
sums. For the nonpolynomial term, we use the Laplace representation
\begin{equation}
\mathcal{R}^{\mathrm{odd}}_{l}(\eta)=4\pi\,2^{l+1}(l+1)!(2\pi)^{l}\sum_{\mathbf{n}\ne0}(-i)^{l}\mathcal{Y}_{lm}(\mathbf{n})\int^{\infty}_{0}du\,\frac{ue^{-\eta u}}{(u^{2}+4\pi^{2}n^{2})^{l+2}}.\label{eq:Rodd-poisson}
\end{equation}
After applying Poisson summation, the odd-power term therefore gives
rise to algebraically decaying contributions through $c^{\mathrm{odd}}_{j,l}$
and $\mathcal{R}^{\mathrm{odd}}_{l}(\eta)$. To evaluate the coefficients
$c_{j,l}$ numerically, we use an accelerated representation of the
corresponding weighted integer sums. The fundamental quantity is
\begin{equation}
\mathcal{A}_{lj}=\sum_{\mathbf{n}\neq0}\frac{\mathcal{Y}_{lm}(\mathbf{n})}{n^{2j+2l+4}},\qquad P=2j+2l+4,
\end{equation}
Using the same finite-part prescription as in the regulated sum, this quantity is evaluated as 
\begin{equation}
\mathcal{A}_{lj}=\sum_{\mathbf{n}\neq0}\frac{\mathcal{Y}_{lm}(\mathbf{n})e^{-n^{P}}}{n^{P}}+4\pi i^{l}\sum_{\mathbf{k}\neq0}Y_{lm}(\hat{\mathbf{k}})\mathcal{I}_{lj}(k)-\frac{\delta_{l0}}{\sqrt{4\pi}}\left[1+\frac{4\pi}{P}\Gamma\left(\frac{3}{P}-1\right)\right],
\end{equation}
where $\mathbf{k}$ is an auxiliary integer vector, and 
\begin{equation}
\mathcal{I}_{lj}(k)=\int^{\infty}_{0}dx\,x^{l+2-P}\left(1-e^{-x^{P}}\right)j_{l}(2\pi kx).
\end{equation}
The radial integral is one-dimensional and can therefore be evaluated numerically.
Equivalently, using 
\begin{equation}
j_{l}(z)=\sqrt{\frac{\pi}{2z}}J_{l+1/2}(z),
\end{equation}
one may write 
\begin{equation}
\mathcal{I}_{lj}(k)=\frac{1}{2\sqrt{k}}\int^{\infty}_{0}dx\,x^{l+\frac{3}{2}-P}\left(1-e^{-x^{P}}\right)J_{l+1/2}(2\pi kx).
\end{equation}
This representation is used to compute the sums entering $c_{j,l}$.
It preserves the finite-part prescription of the regulated sum while
improving the convergence of the original partial-wave expression. 

\subsection{The Case Containing $E_{n}$}

We now extend the preceding algebraic analysis to a non-polynomial
term derived from the relativistic dispersion relation. In addition
to the prefactor arising directly from the two-particle propagator,
it is useful to consider the complementary energy-weighted term $n^{2s}E_{n}$.
This case provides a massive relativistic extension of the odd-power
term because $E_{n}$ reduces to $n$ in the massless limit. When $g(n;\nu)=n^{2s}E_{n}$,
we obtain
\begin{equation}
\mathcal{D}_{s,l}(\eta,\nu)=\left(\sum_{\mathbf{n}\in\mathbb{Z}^{3}}-\mathrm{PV}\int_{\mathbb{R}^{3}}d^{3}\mathbf{n}\right)\left[\frac{n^{2s}E_{n}-\eta^{2s}E_{\eta}}{n^{2}-\eta^{2}}\mathcal{Y}_{lm}(\mathbf{n})\right].\label{eq:D-rel-def}
\end{equation}
Using the algebraic identity 
\begin{equation}
\frac{n^{2s}E_{n}-\eta^{2s}E_{\eta}}{n^{2}-\eta^{2}}=\sum^{s-1}_{j=0}\eta^{2(s-1-j)}n^{2j}E_{n}+\eta^{2s}\frac{1}{E_{n}+E_{\eta}},
\end{equation}
 Eq.~\eqref{eq:D-rel-def} becomes 
\begin{equation}
\mathcal{D}_{s,l}(\eta,\nu)=\sum^{s-1}_{j=0}\eta^{2(s-1-j)}c_{j,l}(\nu)+\eta^{2s}\mathcal{R}_{l}(\eta,\nu).\label{eq:D-rel-decomp}
\end{equation}
The corresponding coefficients and the nonpolynomial term in Eq.~\eqref{eq:D-rel-decomp} are 
\begin{align}
c_{j,l}(\nu) & =\left(\sum_{\mathbf{n}}-\int d^{3}\mathbf{n}\right)n^{2j}\sqrt{n^{2}+\nu^{2}}\,\mathcal{Y}_{lm}(\mathbf{n}),\\
\mathcal{R}_{l}(\eta,\nu) & =\left(\sum_{\mathbf{n}}-\int d^{3}\mathbf{n}\right)\frac{\mathcal{Y}_{lm}(\mathbf{n})}{\sqrt{n^{2}+\nu^{2}}+\sqrt{\eta^{2}+\nu^{2}}}.
\end{align}
The finite-part definitions are obtained by applying the same dimensional regularization to both the sum and the integral: 
\begin{align}
c_{j,l}(\nu) & =\lim\limits_{\epsilon\to0^{+}}\left[\sum_{\mathbf{n}\in\mathbb{Z}^{3}}n^{2j}\sqrt{n^{2}+\nu^{2}}\,\mathcal{Y}_{lm}(\mathbf{n})e^{-\epsilon n^{2}}-\int_{\mathbb{R}^{3}}d^{3}\mathbf{n}\,n^{2j}\sqrt{n^{2}+\nu^{2}}\,\mathcal{Y}_{lm}(\mathbf{n})e^{-\epsilon n^{2}}\right],\\
\mathcal{R}_{l}(\eta,\nu) & =\lim\limits_{\epsilon\to0^{+}}\left[\sum_{\mathbf{n}\in\mathbb{Z}^{3}}\frac{\mathcal{Y}_{lm}(\mathbf{n})e^{-\epsilon n^{2}}}{\sqrt{n^{2}+\nu^{2}}+\sqrt{\eta^{2}+\nu^{2}}}-\int_{\mathbb{R}^{3}}d^{3}\mathbf{n}\,\frac{\mathcal{Y}_{lm}(\mathbf{n})e^{-\epsilon n^{2}}}{\sqrt{n^{2}+\nu^{2}}+\sqrt{\eta^{2}+\nu^{2}}}\right].
\end{align}
The terms involving $c_{j,l}(\nu)$ constitute the part of the decomposition that is polynomial in $\eta^{2}$, whereas $\mathcal{R}_{l}(\eta,\nu)$ contains the non-polynomial dependence on $\eta$.

Applying Poisson summation to the $E_{n}$-weighted coefficients gives
\begin{equation}
c_{j,l}(\nu)=\sum_{\mathbf{n}\ne0}\left(-\frac{\nabla^{2}_{\mathbf{n}}}{4\pi^{2}}\right)^{j}\left\{ (-i)^{l}Y_{lm}(\hat{\mathbf{n}})\left[-\frac{\nu^{l+2}}{\pi n^{2}}K_{l+2}(2\pi\nu n)\right]\right\} .\label{eq:c-rel-poisson-compact}
\end{equation}
After applying the Laplacian to the radial function, this expression may be written as 
\begin{equation}
\begin{aligned}
c_{j,l}(\nu)= & -\pi j!\sum_{\mathbf{n}\neq0}(-i)^{l}Y_{lm}(\hat{\mathbf{n}})(\pi n)^{l}\sum^{j}_{r=0}(-1)^{r}\frac{\Gamma\!\left(j+l+\frac{3}{2}\right)}{\Gamma\!\left(r+l+\frac{3}{2}\right)(j-r)!r!} \\
 & \times(\pi^{2}n^{2})^{r}\left(\frac{\nu}{\pi n}\right)^{j+l+r+2}K_{j+l+r+2}(2\pi\nu n).
\end{aligned}
\label{eq:c-rel-poisson-expanded}
\end{equation}
The expressions in Eqs.~\eqref{eq:c-rel-poisson-compact} and~\eqref{eq:c-rel-poisson-expanded}
are algebraically equivalent. For $\mathcal{R}_{l}(\eta,\nu)$, an
exponential-parameter representation of $(E_{n}+E_{\eta})^{-1}$ yields
\begin{equation}
\mathcal{R}_{l}(\eta,\nu)=4\pi(2\pi)^{l}\nu^{l+2}\sum_{\mathbf{n}\ne0}(-i)^{l}\mathcal{Y}_{lm}(\mathbf{n})\int^{\infty}_{0}du\,\frac{ue^{-u\sqrt{\eta^{2}+\nu^{2}}}}{(u^{2}+4\pi^{2}n^{2})^{\frac{(l+2)}{2}}}K_{l+2}\!\left(\nu\sqrt{u^{2}+4\pi^{2}n^{2}}\right).\label{eq:R-rel-poisson}
\end{equation}
These expressions explicitly show how $\nu$ produces exponential
suppression at large shell radii, governed by factors of the form
$e^{-2\pi\nu n}$ up to algebraic prefactors. At the massless boundary,
the asymptotic form $K_{l+2}(z)\sim2^{l+1}(l+1)!z^{-(l+2)}$ yields
\begin{equation}
\nu^{l+2}K_{l+2}\!\left(\nu\sqrt{u^{2}+4\pi^{2}n^{2}}\right)\longrightarrow2^{l+1}(l+1)!\left(u^{2}+4\pi^{2}n^{2}\right)^{-(l+2)/2},
\end{equation}
and Eq.~\eqref{eq:R-rel-poisson} reduces to Eq.~\eqref{eq:Rodd-poisson}.
Therefore, the decomposition approaches 
\begin{equation}
\mathcal{D}_{s,l}(\eta,\nu)\xrightarrow[\nu\to0]{}\sum^{s-1}_{j=0}\eta^{2(s-1-j)}c^{\mathrm{odd}}_{j,l}+\eta^{2s}\mathcal{R}^{\mathrm{odd}}_{l}(\eta).
\end{equation}
The massive and massless cases share the same algebraic decomposition; the distinction lies in their long-distance behavior, where the exponential suppression for $\nu>0$ is replaced by algebraic decay at $\nu=0$ \cite{ref4,ref5,ref6,ref7,ref9,ref16,ref39,ref18,ref40}.

\subsection{The case containing $1/E_{n}$}

We finally consider the case in which the smooth factor multiplying
the pole contains the factor $1/E_{n}$ introduced in Sec. 2. This factor is
the dimensionless counterpart of the factor $1/(2\omega_{\mathbf{k}})$
that arises after energy integration in the relativistic two-particle
loop.

\begin{equation}
\widetilde{\mathcal{D}}_{s,l}(\eta,\nu)=\left(\sum_{\mathbf{n}\in\mathbb{Z}^{3}}-\mathrm{PV}\int_{\mathbb{R}^{3}}d^{3}\mathbf{n}\right)\left[\frac{n^{2s}/E_{n}-\eta^{2s}/E_{\eta}}{n^{2}-\eta^{2}}\mathcal{Y}_{lm}(\mathbf{n})\right].
\end{equation}
 Using 
\begin{equation}
\frac{1}{E_{n}}-\frac{1}{E_{\eta}}=-\frac{n^{2}-\eta^{2}}{E_{n}E_{\eta}(E_{n}+E_{\eta})},
\end{equation}
one obtains 
\begin{equation}
\frac{n^{2s}/E_{n}-\eta^{2s}/E_{\eta}}{n^{2}-\eta^{2}}=\sum^{s-1}_{j=0}\eta^{2(s-1-j)}\frac{n^{2j}}{E_{n}}-\frac{\eta^{2s}}{E_{\eta}}\frac{1}{E_{n}(E_{n}+E_{\eta})}.
\end{equation}
Thus 
\begin{equation}
\widetilde{\mathcal{D}}_{s,l}(\eta,\nu)=\sum^{s-1}_{j=0}\eta^{2(s-1-j)}\widetilde{c}_{j,l}(\nu)-\frac{\eta^{2s}}{E_{\eta}}\widetilde{\mathcal{R}}_{l}(\eta,\nu)
\end{equation}
where the finite sum is understood to vanish when $s=0$. The corresponding
terms are defined by
\begin{align}
\widetilde{c}_{j,l}(\nu) & =\left(\sum_{\mathbf{n}}-\int d^{3}\mathbf{n}\right)\frac{n^{2j}\mathcal{Y}_{lm}(\mathbf{n})}{E_{n}},\\
\widetilde{\mathcal{R}}_{l}(\eta,\nu) & =\left(\sum_{\mathbf{n}}-\int d^{3}\mathbf{n}\right)\frac{\mathcal{Y}_{lm}(\mathbf{n})}{E_{n}(E_{n}+E_{\eta})}.
\end{align}
The finite-part definitions are 
\begin{align}
\widetilde{c}_{j,l}(\nu) & =\lim\limits_{\epsilon\to0^{+}}\left[\sum_{\mathbf{n}\in\mathbb{Z}^{3}}\frac{n^{2j}\mathcal{Y}_{lm}(\mathbf{n})}{E_{n}}e^{-\epsilon n^{2}}-\int_{\mathbb{R}^{3}}d^{3}\mathbf{n}\,\frac{n^{2j}\mathcal{Y}_{lm}(\mathbf{n})}{E_{n}}e^{-\epsilon n^{2}}\right],\\
\widetilde{\mathcal{R}}_{l}(\eta,\nu) & =\lim\limits_{\epsilon\to0^{+}}\left[\sum_{\mathbf{n}\in\mathbb{Z}^{3}}\frac{\mathcal{Y}_{lm}(\mathbf{n})}{E_{n}(E_{n}+E_{\eta})}e^{-\epsilon n^{2}}-\int_{\mathbb{R}^{3}}d^{3}\mathbf{n}\,\frac{\mathcal{Y}_{lm}(\mathbf{n})}{E_{n}(E_{n}+E_{\eta})}e^{-\epsilon n^{2}}\right].
\end{align}
Applying Poisson summation using the same convention as in the preceding subsection, the weighted coefficients take the form 
\begin{equation}
\widetilde{c}_{j,l}(\nu)=\sum_{\mathbf{n}\neq0}\left(-\frac{\nabla^{2}_{\mathbf{n}}}{4\pi^{2}}\right)^{j}\left[(-i)^{l}\mathcal{Y}_{lm}(\mathbf{n})\frac{2\nu^{l+1}}{n^{l+1}}K_{l+1}(2\pi\nu n)\right].
\end{equation}
 After applying the Laplacian to the radial function, this expression may be written as 
\begin{equation}
\widetilde{c}_{j,l}(\nu)=\sum_{\mathbf{n}\neq0}(-i)^{l}\mathcal{Y}_{lm}(\mathbf{n})\sum^{j}_{r=0}\binom{j}{r}(-\nu^{2})^{j-r}\frac{2\pi^{\frac{1}{2}-r}\nu^{r+l+1}}{\Gamma\left(\frac{1}{2}-r\right)n^{r+l+1}}K_{l+r+1}(2\pi\nu n).
\end{equation}

For $\widetilde{\mathcal{R}}_{l}(\eta,\nu)$, we introduce the following Laplace representation:
\begin{equation}
\frac{1}{E_{n}(E_{n}+E_{\eta})}=\int^{\infty}_{0}du\,e^{-uE_{\eta}}\frac{e^{-uE_{n}}}{E_{n}}.
\end{equation}
Applying Poisson summation and then performing the angular integration yield
\begin{equation}
\widetilde{\mathcal{R}}_{l}(\eta,\nu)=4\pi(2\pi)^{l}\nu^{l+1}\sum_{\mathbf{n}\ne0}(-i)^{l}\mathcal{Y}_{lm}(\mathbf{n})\int^{\infty}_{0}du\,\frac{e^{-u\sqrt{\eta^{2}+\nu^{2}}}}{\left(u^{2}+4\pi^{2}n^{2}\right)^{(l+1)/2}}K_{l+1}\!\left(\nu\sqrt{u^{2}+4\pi^{2}n^{2}}\right).
\end{equation}

For finite $\nu$, the modified Bessel functions yield exponential suppression governed by factors of the form $e^{-2\pi\nu n}$, up to algebraic prefactors. At the massless boundary, the small-argument behavior 
\begin{equation}
K_{l+1}(z)\sim2^{l}l!z^{-(l+1)}
\end{equation}
gives 
\begin{equation}
\nu^{l+1}K_{l+1}\!\left(\nu\sqrt{u^{2}+4\pi^{2}n^{2}}\right)\longrightarrow2^{l}l!\left(u^{2}+4\pi^{2}n^{2}\right)^{-(l+1)/2}.
\end{equation}
Thus, for $\nu>0$, the contribution containing the factor $1/E_{n}$
is exponentially suppressed by $e^{-2\pi\nu n}$. In the $\nu\to0$
limit, this exponential suppression is replaced by algebraic decay.
The terms containing the factors $E_{n}$ and $1/E_{n}$ arise from the
same algebraic decomposition of the on-shell subtraction; however, they
lead to different Bessel indices and distinct power laws in the massless
limit.

The factor $1/E_{n}$ accompanies the two-particle on-shell pole in
the relativistic propagator. In the traditional Lüscher derivation,
the pole contribution that generates the power-law finite-volume
dependence is retained, whereas the difference introduced by the subtraction
$1/E_{n}\to1/E_{\eta}$ is treated as an exponentially suppressed
finite-volume correction \cite{ref13,ref15,ref25}.

\subsection{The S-wave near the formal massless boundary}

Near a two-hadron threshold, the centrifugal barrier suppresses partial-wave amplitudes with $l>0$, whereas the $S$-wave amplitude ($l=0$) is not subject to such suppression. When permitted by the quantum numbers, the $S$-wave therefore generally provides the leading contribution to the low-energy expansion \cite{ref25,ref26}. Motivated by this hierarchy, we restrict the analysis in this subsection to $l=0$ and compare the behavior of the cases involving $E_{n}$ and $1/E_{n}$ as $\nu\to0^{+}$. This specialization does not affect the general $l$-dependent expressions derived above. For fixed $\eta>0$, the massless relations derived above reduce in the $S$-wave case to 
\begin{equation}
\begin{aligned}\lim_{\nu\to0^{+}}\mathcal{D}_{s,0}(\eta,\nu) & =\mathcal{D}^{\mathrm{odd}}_{s,0}(\eta), &  & s\geq0,\\
\lim_{\nu\to0^{+}}\widetilde{\mathcal{D}}_{s,0}(\eta,\nu) & =\mathcal{D}^{\mathrm{odd}}_{s-1,0}(\eta), &  & s\geq1.
\end{aligned}
\end{equation}

The first relation follows from $n^{2s}E_{n}\to n^{2s+1}$, whereas
the second follows from $n^{2s}/E_{n}\to n^{2s-1}$. Thus, whenever
the endpoint exists, the two cases reproduce the corresponding odd-power
structures. The case $s=0$ is excluded from the second relation.
To distinguish the manner in which these endpoints are approached, we next expand
the complete $S$-wave expressions at fixed $\eta>0$.

For the case involving $E_{n}$, the endpoint is finite for every
$s$. For the case involving $1/E_{n}$, one finds 

\begin{equation}
\widetilde{\mathcal{D}}_{s,0}(\eta,\nu)=\begin{cases}
-\dfrac{\mathcal{Y}_{00}}{\eta^{2}\nu}+O(1), & s=0,\\[8pt]
\mathcal{D}^{\mathrm{odd}}_{s-1,0}(\eta)+O(\nu^{2}), & s\geq1,
\end{cases}\qquad\nu\to0^{+}.\label{eq:inverse-En-S-wave-complete-small-nu}
\end{equation}
 The $s=0,\ l=0$ case is qualitatively different: $\widetilde{\mathcal{D}}_{0,0}(\eta,\nu)$ has no finite limit as $\nu\to0^{+}$.

This exceptional behavior arises from the discrete zero-momentum term. Since $\mathcal{Y}_{00}(\mathbf{0})=\mathcal{Y}_{00}$, 
\begin{equation}
\left.\frac{1/E_{n}-1/E_{\eta}}{n^{2}-\eta^{2}}\mathcal{Y}_{00}(\mathbf{n})\right|_{\mathbf{n}=\mathbf{0}}=-\frac{\mathcal{Y}_{00}}{\nu E_{\eta}(\nu+E_{\eta})}=-\frac{\mathcal{Y}_{00}}{\eta^{2}\nu}+O(1).
\end{equation}

This zero-momentum divergence is absent for $l>0$, since $\mathcal{Y}_{lm}(\mathbf{0})=0$.

The corresponding continuum integrand is locally integrable at the origin. Indeed, for $\nu=0$, 
\begin{equation}
\frac{1/n-1/\eta}{n^{2}-\eta^{2}}\mathcal{Y}_{00}=-\frac{\mathcal{Y}_{00}}{\eta n(n+\eta)}\sim-\frac{\mathcal{Y}_{00}}{\eta^{2}n},\qquad n\to0.
\end{equation}

Together with the three-dimensional measure, this behavior yields a radial contribution proportional to $n\,dn$ near the origin and is therefore locally integrable. Consequently, the continuum integral does not generate a corresponding $1/\nu$ divergence, and the singular behavior of $\widetilde{\mathcal{D}}_{0,0}$ arises entirely from the discrete zero-momentum mode.

\subsection{On-shell subtraction below threshold}

The preceding derivations were formulated for real $\eta$, corresponding to a real relative momentum in the elastic-scattering region. Below the two-particle threshold, the relative momentum in the center-of-mass frame becomes purely imaginary \cite{ref13,ref25}. We therefore write
\begin{equation}
\eta=i\kappa,\qquad\kappa\geq0.
\end{equation}
where $\kappa=0$ corresponds to the two-particle threshold, while $\kappa>0$
describes the below-threshold region. For real $\eta$, the denominator
$n^{2}-\eta^{2}$ may vanish within the real integration domain; therefore,
a principal-value prescription is required. Below the threshold, 

\begin{equation}
n^{2}-\eta^{2}=n^{2}+\kappa^{2},
\end{equation}
which has no zeros in the real momentum domain for $\kappa>0$. Consequently, no principal-value prescription is required for the below-threshold expressions. The polynomial identity underlying the even-power case remains valid below threshold. Under the same dimensional-regularization finite-part prescription, the corresponding sum–integral term therefore satisfies
\begin{equation}
\mathcal{D}^{\mathrm{even}}_{s,l}(i\kappa)=0.
\end{equation}
For the odd-power case, the decomposition takes the following form:

\begin{equation}
\mathcal{D}^{\mathrm{odd}}_{s,l}(i\kappa)=\sum^{s-1}_{j=0}(i\kappa)^{2(s-1-j)}c^{\mathrm{odd}}_{j,l}+(i\kappa)^{2s}\mathcal{R}^{\mathrm{odd}}_{l}(i\kappa).
\end{equation}
The coefficients $c^{\mathrm{odd}}_{j,l}$ remain unchanged, while
the dependence on $\kappa$ enters through the explicit powers of $i\kappa$
and $\mathcal{R}^{\mathrm{odd}}_{l}(i\kappa)$.

For the relativistic case, we define
\begin{equation}
E_{\eta}=\sqrt{\eta^{2}+\nu^{2}}.
\end{equation}
On the branch analytically continued from positive real values of $\eta$,
\begin{equation}
E_{i\kappa}=\sqrt{\nu^{2}-\kappa^{2}}>0,\qquad0\leq\kappa<\nu.
\end{equation}
In this domain,
\begin{equation}
\mathcal{D}_{s,l}(i\kappa,\nu)=\sum^{s-1}_{j=0}(i\kappa)^{2(s-1-j)}c_{j,l}(\nu)+(i\kappa)^{2s}\mathcal{R}_{l}(i\kappa,\nu).
\end{equation}
The coefficients $c_{j,l}(\nu)$ remain unchanged, whereas the below-threshold
dependence enters through the explicit powers of $\kappa$, $E_{i\kappa}$,
and $\mathcal{R}_{l}(i\kappa,\nu)$.

For the case involving $1/E_{n}$, the continued decomposition is
\begin{equation}
\widetilde{\mathcal{D}}_{s,l}(i\kappa,\nu)=\sum^{s-1}_{j=0}(i\kappa)^{2(s-1-j)}\widetilde{c}_{j,l}(\nu)-\frac{(i\kappa)^{2s}}{E_{i\kappa}}\widetilde{\mathcal{R}}_{l}(i\kappa,\nu).
\end{equation}
The coefficients $\widetilde{c}_{j,l}(\nu)$ also remain unchanged.
In this case, the factor $1/E_{i\kappa}$ becomes singular as $\kappa\to\nu$,
consistent with the domain specified above. Thus, all four decompositions
retain their algebraic structure below the threshold, with unchanged
coefficients.

\section{Numerical results }

The analytic expressions derived in Sec. 3 are evaluated numerically
to examine the dependence of the on-shell subtraction terms on the mass
parameter $\nu$, the external two-hadron momentum, and the partial-wave
structure. The even-power case does not contribute under the prescription
adopted here, while the standard odd-power result is recovered in
the massless limit of the case involving $E_{n}$. We therefore focus on
the cases involving $E_{n}$ and $1/E_{n}$.

The quantities $\mathcal{D}_{s,l}$ and $\widetilde{\mathcal{D}}_{s,l}$
describe the sum$-$integral differences associated with the on-shell
subtraction. Their contributions to complete hadronic observables
also depend on the interaction, the cubic-group projection,
and the finite-volume prescription, which will be discussed in Sec.
5. In the numerical examples, we evaluate the $m=0$ component. The
other components related by cubic symmetry are determined by the relations
summarized in Appendix A.

\subsection{Dependence on the dimensionless mass parameter $\nu$}

\begin{figure}[!htbp]
\centering \includegraphics[width=0.86\textwidth]{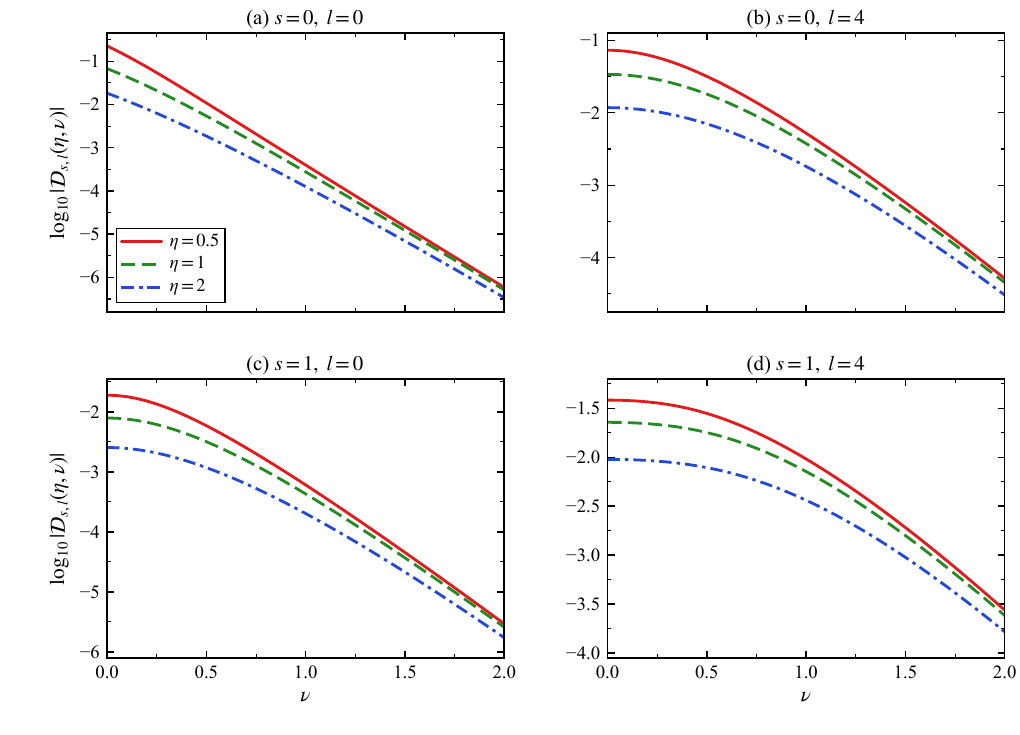}
\caption{(color online) Base-10 logarithm of $|\mathcal{D}_{s,l}(\eta,\nu)|$ as a function of $\nu$ for $s=0,1$, $l=0,4$, and $\eta=0.5,1.0,2.0$ over $0\protect\leq\nu\protect\leq2$. The upper and lower rows correspond to $s=0$ and $s=1$, respectively, whereas the left and right columns correspond to $l=0$ and $l=4$, respectively. Near $\nu=0$, the $s=0,\ l=0$ curves exhibit an approximately linear decrease, whereas the $l=4$ curves are initially flatter and subsequently develop stronger curvature.}
\label{fig:completeD-mass-dependence} 
\end{figure}

\begin{figure}[!htbp]
\centering \includegraphics[width=0.86\textwidth]{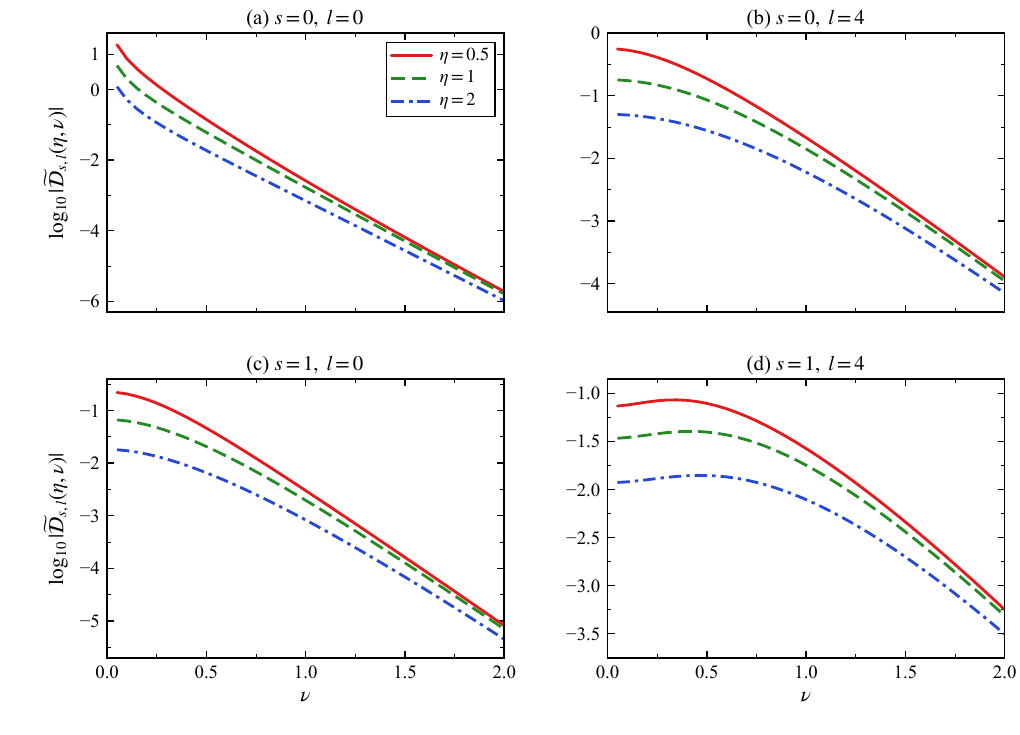}
\caption{(color online) Base-10 logarithm of $|\widetilde{\mathcal{D}}_{s,l}(\eta,\nu)|$ as a function of $\nu$ for $s=0,1$, $l=0,4$, and $\eta=0.5,1.0,2.0$ over the range $0<\nu\protect\leq2$. The upper and lower rows correspond to $s=0$ and $s=1$, respectively, whereas the left and right columns correspond to $l=0$ and $l=4$, respectively.}
\label{fig:inverse-energy-mass-dependence} 
\end{figure}

For $\nu>0$, the finite-volume corrections are expressed in terms
of modified Bessel functions and contain exponential factors $e^{-2\pi\nu|\mathbf{n}|}=e^{-\mu L|\mathbf{n}|}$,
whereas, in the massless limit, they exhibit power-law finite-volume dependence
\cite{ref4,ref5,ref6,ref7}. Figures~\ref{fig:completeD-mass-dependence}
and \ref{fig:inverse-energy-mass-dependence} show the dependence
of the complete sum--integral differences on $\nu$ for the $E_{n}$
and $1/E_{n}$ cases, respectively.

Figure~\ref{fig:completeD-mass-dependence} shows the mass dependence
of the finite-volume correction in the case involving $E_{n}$.
In all four panels, $\left|\mathcal{D}_{s,l}(\eta,\nu)\right|$ decreases
with increasing $\nu$. For sufficiently large $\nu$, this behavior is
consistent with the exponential suppression induced by a nonzero
mass parameter. After application of the Poisson summation formula,
the modified Bessel functions of the second kind contain the asymptotic
factor $e^{-2\pi\nu|\mathbf{n}|}$ multiplied by algebraic prefactors.
At small $\nu$, the decrease is slower than in the large-$\nu$ regime,
and the approach to the massless limit is governed by the small-$\nu$
expansion derived in Sec. 3.5. In particular, the $s=0,l=0$ curves
approach finite endpoints as $\nu\to0^{+}$.

Figure~\ref{fig:inverse-energy-mass-dependence} shows the mass dependence
of the quantities containing $1/E_{n}$. At moderate and large values
of $\nu$, the magnitudes $\lvert\widetilde{\mathcal{D}}_{s,l}(\eta,\nu)\rvert$
generally decrease as $\nu$ increases. As in the $E_{n}$ case, this
behavior is consistent with the large-argument asymptotic form of the
modified Bessel functions obtained in Sec.~3, which yields exponential
suppression at nonzero $\mu L$. At small $\nu$, however, the decrease
in $\lvert\widetilde{\mathcal{D}}_{s,l}(\eta,\nu)\rvert$ becomes less
rapid. The $\widetilde{\mathcal{D}}_{0,0}$ component requires separate
consideration near $\nu=0$.

In Fig.~\ref{fig:inverse-energy-mass-dependence}(a), $\lvert\widetilde{\mathcal{D}}_{0,0}(\eta,\nu)\rvert$
exceeds unity at small $\nu$, indicating that the finite-volume term
associated with long-range propagation becomes numerically large in
this region. Moreover, $\widetilde{\mathcal{D}}_{0,0}$ diverges at
$\nu=0$. As derived in Eq.~\eqref{eq:inverse-En-S-wave-complete-small-nu},
\begin{equation}
\widetilde{\mathcal{D}}_{0,0}(\eta,\nu)=-\frac{\mathcal{Y}_{00}}{\eta^{2}\nu}+O(1),\qquad\nu\to0^{+}.
\end{equation}
Accordingly, $\lvert\widetilde{\mathcal{D}}_{0,0}(\eta,\nu)\rvert$
grows without bound as $\nu\to0^{+}$. This behavior reflects the
infrared sensitivity associated with long-range propagation. Finite-volume
calculations involving massless fields therefore require an explicit
infrared prescription, including a specified treatment of the zero-momentum
mode \cite{ref6,ref9,ref16,ref17}.

The factor $1/\eta^{2}$ determines the relative separation among the
three curves, with the magnitude being largest at the smallest value
of $\eta$. The $1/\nu$ term arises from the discrete zero-momentum
mode in the finite-volume sum. By contrast, the corresponding infinite-volume
integral remains locally integrable as the loop momentum vanishes and
therefore does not cancel this contribution in the sum$-$integral difference.

Taken together, Figs.~\ref{fig:completeD-mass-dependence} and \ref{fig:inverse-energy-mass-dependence}
show that the structures containing $E_{n}$ and $1/E_{n}$ exhibit
the same type of exponential suppression at larger values of $\nu$,
whereas their behavior near $\nu=0$ can differ substantially. The factor
$1/E_{n}$ follows directly from the relativistic two-particle propagator,
whereas the $E_{n}$ structure provides a finite-mass generalization
whose massless limit is related to the standard odd-power case. Their
distinct dependences on the loop energy therefore yield different
finite-volume contributions to the on-shell subtraction discrepancy,
even at fixed $s$, $l$, and $\eta$.

\subsection{Dependence on the on-shell relative-momentum parameter $\eta$}

\begin{figure}[!htbp]
\centering \includegraphics[width=0.70\textwidth]{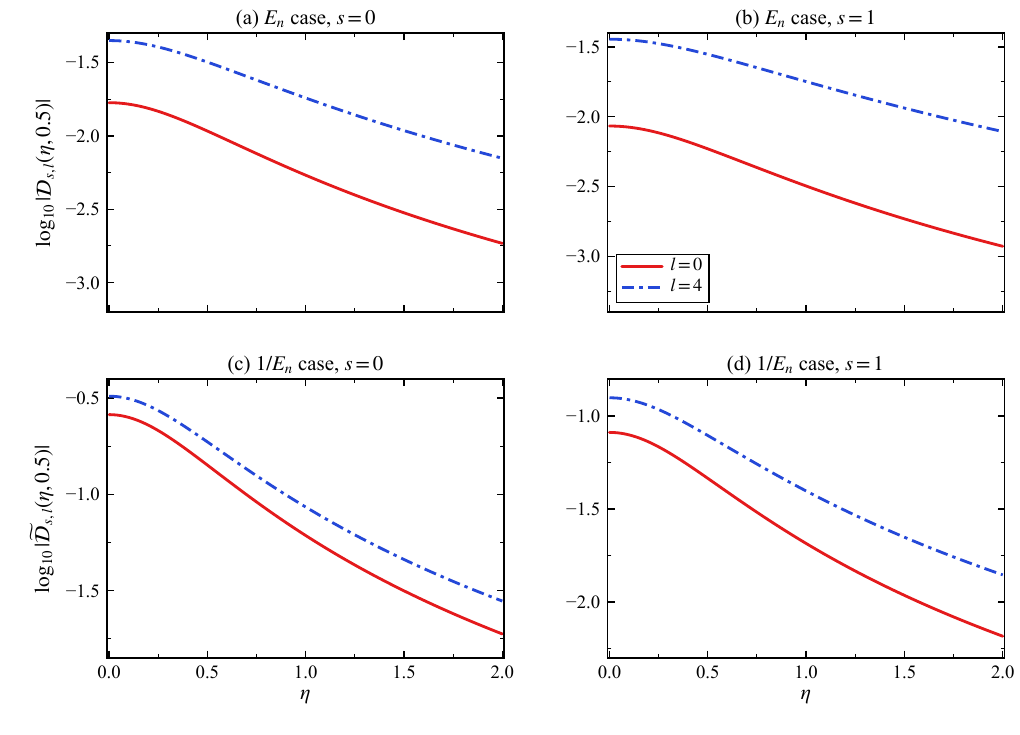}
\caption{(color online) Base-10 logarithms of $\lvert\mathcal{D}_{s,l}(\eta,0.5)\rvert$
and $|\widetilde{\mathcal{D}}_{s,l}(\eta,0.5)|$ as functions of the
on-shell relative-momentum parameter $\eta$. The upper panels show
$\mathcal{D}_{s,l}$ for (a) $s=0$ and (b) $s=1$, whereas the lower
panels show $\widetilde{\mathcal{D}}_{s,l}$ for (c) $s=0$ and (d)
$s=1$. The curves correspond to $l=0$ and $l=4$, as indicated in
the legend, over the range $0\protect\leq\eta\protect\leq2$.}
\label{fig:combined-momentum-dependence} 
\end{figure}

We next examine the dependence on the external on-shell relative momentum by fixing $\nu=0.5$ and continuously varying $\eta$. Fig.~\ref{fig:combined-momentum-dependence}
shows $\lvert\mathcal{D}_{s,l}(\eta,0.5)\rvert$ and $|\widetilde{\mathcal{D}}_{s,l}(\eta,0.5)|$
for $0\leq\eta\leq2$. 

\FloatBarrier

Both $\lvert\mathcal{D}_{s,l}(\eta,0.5)\rvert$ and $\lvert\widetilde{\mathcal{D}}_{s,l}(\eta,0.5)\rvert$
decrease as $\eta$ increases over the displayed range. In the cases
shown, as $\eta$ increases from 0 to 2, the $l=0$ components exhibit a
larger decrease in their base-10 logarithmic magnitudes than the
corresponding $l=4$ components. This comparison is limited to the
parameter values and momentum range shown in
Fig.~\ref{fig:combined-momentum-dependence}.

\subsection{Cancellation in the on-shell subtraction discrepancy}

\begin{figure}[!htbp]
\centering \includegraphics[width=0.70\textwidth]{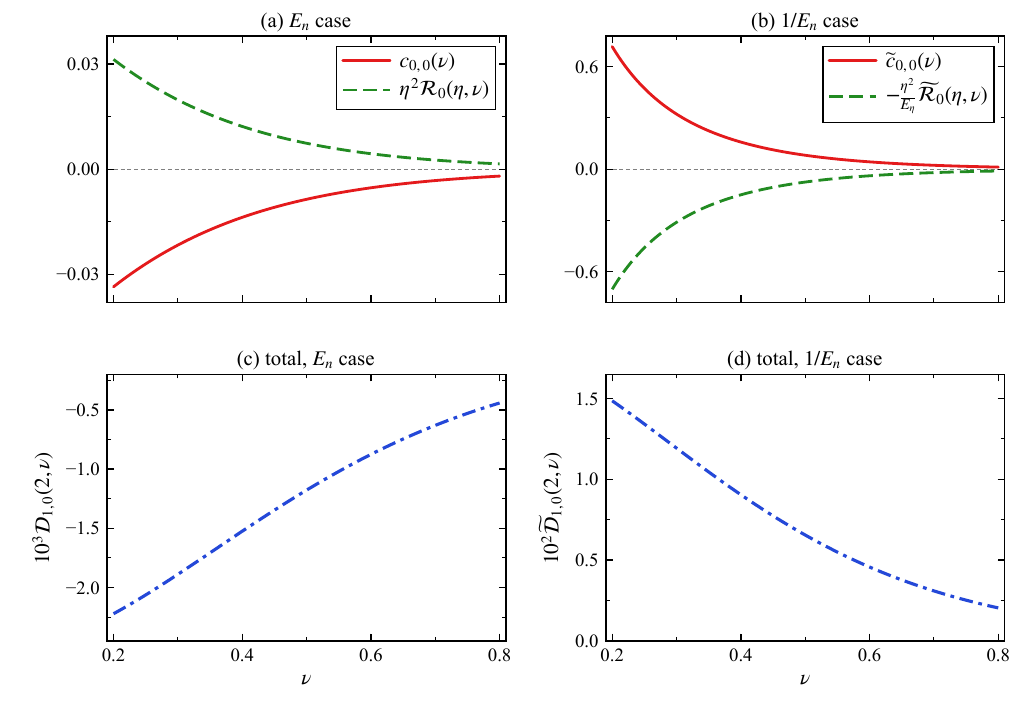}
\caption{(Color online) Cancellation between terms in the decompositions of $\mathcal{D}_{1,0}(2,\nu)$
and $\widetilde{\mathcal{D}}_{1,0}(2,\nu)$ for $0.2\protect\leq\nu\protect\leq0.8$.
The upper-left panel shows $c_{0,0}(\nu)$ and $\eta^{2}\mathcal{R}_{0}(\eta,\nu)$,
whereas the upper-right panel shows $\widetilde{c}_{0,0}(\nu)$ and
$-\eta^{2}\widetilde{\mathcal{R}}_{0}(\eta,\nu)/E_{\eta}$, with $\eta=2$.
The lower panels show the corresponding complete sum$-$integral differences
using enlarged vertical scales.}
\label{fig:signed-decomposition} 
\end{figure}

The preceding figures display only the absolute values of the complete
results and therefore do not reveal cancellations among the terms
in the decompositions. To examine their signed combination, Figure~\ref{fig:signed-decomposition}
presents the results for $s=1$, $l=0$, and $\eta=2$.

\FloatBarrier

For both $\mathcal{D}_{1,0}(2,\nu)$ and $\widetilde{\mathcal{D}}_{1,0}(2,\nu)$,
the two terms have opposite signs throughout the displayed interval
and therefore partially cancel. Consequently, the full sum$-$integral
differences are substantially smaller in magnitude than the individual
terms. Across the four figures, increasing the mass parameter generally
reduces the finite-volume sum$-$integral discrepancies. However, the
continuous momentum scan shows that their detailed dependence also
varies with $s$, the relativistic factor, and the cubic-angular component.
The numerical sums were evaluated using progressively larger truncation
ranges, and the results shown above were stable at the resolution
of the figures. The smaller totals in Fig.~\ref{fig:signed-decomposition}
were also verified using larger truncation ranges and higher numerical
precision; their signs and displayed orders of magnitude remained
stable.

These results numerically illustrate the analytic structure derived
in Sec.~3. They characterize the kinematic sum$-$integral functions
associated with the on-shell subtraction. Their role in finite-volume
hadronic observables depends on the process-dependent structures
discussed in Sec.~5.

\FloatBarrier

\section{Physical implications and applications}

The quantities derived in the preceding sections characterize the
finite-volume corrections arising from the on-shell subtraction of
momentum-dependent factors in two-particle loop functions. They contribute
to observables such as finite-volume energy shifts \cite{ref24},
hadron masses \cite{ref10,ref11,ref12,ref19,ref21,ref22,ref23,ref37},
and matrix elements \cite{ref20,ref29} through kinematic sum$-$integral
differences associated with the momentum dependence. The physical
significance of these corrections lies in their role in determining the
validity of the on-shell subtraction commonly employed in finite-volume
scattering formalisms. In particular, the results obtained in this work
provide the kinematic information needed to assess the accuracy of the
on-shell subtraction in a given finite-volume calculation.

\subsection{Implications for the on-shell subtraction }

In Lüscher-type finite-volume derivations, the two-particle pole is
isolated before regular momentum-dependent factors in the two-particle
loop are evaluated at their on-shell values. The pole term generates
the power-law finite-volume dependence that enters the quantization condition.
For short-range interactions, if the pole-subtracted integrand is
analytic in the momentum domain relevant to the sum$-$integral relation
and contains no additional low-energy singularities, the corresponding
finite-volume correction is exponentially suppressed \cite{ref31,ref13,ref25,ref15}.
Under these assumptions, the on-shell subtraction reproduces the standard
quantization condition up to exponentially suppressed corrections.

The analysis in Sec.~3 evaluates the sum$-$integral difference generated
by this subtraction for the radial structures considered in this work.
For the polynomial dependence $g(n)=n^{2s}$, the associated difference
vanishes under the common dimensional-regularization finite-part
prescription adopted for both the sum and the integral. The subtraction $n^{2s}\to\eta^{2s}$
therefore generates no additional contribution for this polynomial
class under the stated prescription. By contrast, the odd-power momentum dependence
and the cases involving $E_{n}$ and $1/E_{n}$ yield nonzero sum$-$integral
differences. The polynomial and non-polynomial terms derived in Sec.~3
are algebraic components of a single decomposition, and their sum
defines the complete contribution associated with the on-shell subtraction.
For $\nu>0$, the cases involving $E_{n}$ and $1/E_{n}$ contain
exponentially suppressed terms of the form $e^{-2\pi\nu|\mathbf{n}|}=e^{-\mu L|\mathbf{n}|}$,
where $\mu$ is the mass parameter entering the corresponding momentum-dependent
structure. Odd-power momentum dependence and massless propagation can
generate algebraic finite-volume terms and consequently contribute
at power-law order \cite{ref40}.

The $s=0,\ l=0$ component in the case involving $1/E_{n}$ requires
separate treatment. When the zero-momentum mode is included in the finite-volume
sum, its leading behavior is proportional to $1/\nu$, and the corresponding
sum$-$integral difference diverges as $\nu\to0^{+}$, as shown in
Eq.~\eqref{eq:inverse-En-S-wave-complete-small-nu}. The definition
of this component in the massless limit therefore depends explicitly
on the prescribed treatment of the zero-momentum mode. The accuracy
of the on-shell subtraction is governed by the interaction range,
the relevant mass scale, the finite-volume mode prescription, and
the required precision of the calculation. Its contribution to a specified
hadronic observable is obtained after restoring the process-dependent coefficients,
angular projections, and powers of $2\pi/L$.

\subsection{Incorporation into finite-volume calculations}

The finite-volume functions derived in this work describe the kinematic
correction induced by the on-shell subtraction. To apply these results
to a physical observable, they must be combined with the process-dependent
structures that appear in the corresponding finite-volume loop calculation
\cite{ref25,ref34,ref35,ref36}. For a generic observable $\mathcal{O}$,
the corresponding finite-volume contribution can be written schematically as
\begin{equation}
\Delta_{L}\mathcal{O}=\left[\frac{1}{L^{3}}\sum_{\mathbf{k}}-\int\frac{d^{3}\mathbf{k}}{(2\pi)^{3}}\right]\mathcal{K}_{\mathcal{O}}(\mathbf{k};q_{\mathrm{on}},\ldots),
\end{equation}
where $\mathcal{K}_{\mathcal{O}}$ denotes the momentum-dependent factors obtained after the relevant energy integrations. It contains the process-dependent momentum structures and coefficients associated with the specific observable, whereas the finite-volume dependence arising from the on-shell subtraction is described by the sum$-$integral differences studied in this work.

For a calculation involving several loop topologies, channels, or propagator structures, this contribution may be organized schematically as 
\begin{equation}
\left.\Delta_{L}\mathcal{O}\right|_{\mathrm{sub}}=\sum_{\alpha}\sum_{s,l}A^{\mathcal{O}}_{\alpha;s,l}(L)\,\mathcal{D}_{s,l}(\eta_{\alpha},\nu_{\alpha})+\sum_{\alpha}\sum_{s,l}\widetilde{A}^{\mathcal{O}}_{\alpha;s,l}(L)\,\widetilde{\mathcal{D}}_{s,l}(\eta_{\alpha},\nu_{\alpha})+\cdots.
\end{equation}
Here $\alpha$ labels the relevant loop topology, two-particle channel,
or propagator structure. The functions $\mathcal{D}_{s,l}$ and $\widetilde{\mathcal{D}}_{s,l}$
describe the structures containing $E_{n}$ and $1/E_{n}$, respectively.
A given hadronic calculation may contain both classes, and their signed
sum determines the complete contribution associated with the on-shell
subtraction.

The coefficients $A^{\mathcal{O}}_{\alpha;s,l}(L)$ and $\widetilde{A}^{\mathcal{O}}_{\alpha;s,l}(L)$
contain the process-dependent couplings, normalization factors, channel
and spin structures, angular projections, and powers of $2\pi/L$
required to restore the physical dimensions. The parameters $\eta_{\alpha}$
and $\nu_{\alpha}$ are fixed by the on-shell kinematics and propagator
structure, with $\nu_{\alpha}=\frac{\mu_{\alpha}L}{2\pi}$, where $\mu_{\alpha}$
denotes the mass parameter entering the corresponding momentum-dependent
factor. The angular-projection and cubic-symmetry constraints are understood
throughout this work, and the magnetic indices are omitted for simplicity.

The numerical results in Sec.~4 therefore characterize the magnitudes
and parameter dependence of the individual kinematic building blocks.
For a specified observable, the physical contribution is obtained
from the complete signed combination in the equation above. This organization
allows the same functions to be used for different hadronic processes
while retaining the interaction and channel dependence in the coefficients.

\subsection{Long-range interactions and finite-volume QCD+QED}

Long-range electromagnetic interactions provide a setting in which
the massless limits derived above can contribute to a hadronic finite-volume
calculation. In the standard Lüscher analysis of short-range interactions,
the finite-volume contribution associated with a smooth, pole-subtracted
loop integrand is exponentially suppressed. By contrast, massless-photon
propagation generally produces algebraic finite-volume dependence, whose
leading power and coefficient are determined by the observable and the
chosen finite-volume QED formulation \cite{ref6,ref7,ref8,ref9,ref40}.
Consequently, the contribution from the on-shell subtraction must be
assessed whenever an electromagnetic loop, after loop-energy integration
and on-shell subtraction, contains one of the massless momentum dependences
analyzed in Sec.~3.

In Sec.~2, the parameter $\nu=m_{h}L/(2\pi)$ is defined in terms of the
particle mass entering the relativistic factor
$E_{n}=\sqrt{n^{2}+\nu^{2}}$ in the two-hadron loop. After loop-energy
integration, electromagnetic loop diagrams can generate momentum factors
associated with both massive hadrons and massless photons
\cite{ref20,ref29,ref37,ref40}.

A term can be expressed in terms of $\mathcal{D}_{s,l}$ or
$\widetilde{\mathcal{D}}_{s,l}$ when its pole-subtracted form has the same
on-shell denominator, radial dependence, angular projection, and
finite-volume mode prescription as the corresponding expression in Sec.~3.
In the notation of Sec.~5.2, the index $\alpha$ distinguishes the propagator
structures and their associated values of $\eta_{\alpha}$ and
$\nu_{\alpha}$. For a term that satisfies these matching conditions, a
positive value of $\nu_{\alpha}$ identifies a massive factor, whereas
$\nu_{\alpha}=0$ identifies a massless factor.

The finite-volume sums evaluated in Secs.~3 and 4 include the discrete
zero-momentum mode. Equation~\eqref{eq:inverse-En-S-wave-complete-small-nu}
shows that this mode generates the leading $1/\nu$ behavior of
$\widetilde{\mathcal{D}}_{0,0}$ as $\nu\to0^{+}$. This behavior quantifies
the sensitivity of the $s=0,\ l=0$ component to the infrared definition
of the finite-volume theory. Several finite-volume QED formulations treat
this infrared sector differently, including photon zero-mode subtraction
\cite{ref6,ref7,ref8,ref9}, a photon-mass regulator \cite{ref16}, and
$C^{\star}$ spatial boundary conditions \cite{ref17,ref30,ref38}. The
sum$-$integral difference appropriate to a given formulation must employ
the same photon momentum modes and infrared regulator as the complete
finite-volume calculation. In a zero-mode-subtracted formulation, the
photon momentum sum is restricted to the set of modes specified by that
formulation.

Once the matched functions have been defined using the selected
finite-volume prescription, they enter the multi-structure expansion in
Sec.~5.2, together with the electromagnetic couplings, hadronic interaction
terms, channel and spin factors, and the required powers of $2\pi/L$. The
signed sum of all matched terms yields the part of the selected finite-volume
quantity generated by the on-shell subtraction. Its magnitude relative to
the other electromagnetic finite-volume contributions is determined by the
process-dependent coefficients, kinematics, and infrared prescription used
for that observable \cite{ref20,ref29,ref37}.

\section{Summary and outlook}

We have studied the finite-volume corrections generated by the on-shell
subtraction of regular momentum-dependent factors multiplying a two-particle
pole in finite-volume loop calculations. The four radial structures
$n^{2s}$, $n^{2s+1}$, $n^{2s}E_{n}$, and $n^{2s}/E_{n}$ were analyzed
using a partial-wave expansion, a Cauchy principal-value prescription
for the real on-shell pole, and a common dimensionally regularized
finite part for the sums and integrals. The factor $1/E_{n}$ carries
the loop-momentum dependence of the relativistic two-particle propagator
discussed in Sec. 2, whereas the $E_{n}$ structure provides a finite-mass
extension that approaches the odd-power case. Within the adopted
dimensional-regularization prescription, the even-power sum$-$integral
term vanishes. The odd-power case and the cases containing $E_{n}$
and $1/E_{n}$ each separate into a finite polynomial in the squared
on-shell relative momentum and a term containing non-polynomial
dependence on the on-shell momentum.

For the relativistic $E_{n}$ and $1/E_{n}$ structures, application
of the Poisson summation formula yields representations involving modified
Bessel functions, whereas the odd-power case yields algebraically
decaying terms. For $\nu>0$, the terms associated with nonzero integer
vectors contain exponential factors of the form
$e^{-2\pi\nu|\mathbf{n}|}=e^{-\mu L|\mathbf{n}|}$ and are suppressed
at large $\mu L$. In the limit $\nu\to0^{+}$, these terms exhibit
enhanced finite-volume dependence. The case containing $E_{n}$
approaches the odd-power result with the same index $s$. By contrast,
when the discrete zero-momentum mode is retained in the case containing
$1/E_{n}$, the $s=0,\ l=0$ component has the leading behavior
$-\mathcal{Y}_{00}/(\eta^{2}\nu)$ as $\nu\to0^{+}$. A massless
implementation of this component therefore requires an explicit
prescription for the zero-momentum mode.

For the massive structures considered here, the correction associated
with the on-shell subtraction is exponentially suppressed at large
$\mu L$ and may be omitted if it lies beyond the desired accuracy.
If massless propagation or an additional infrared singularity eliminates
this suppression, the correction must instead be included in the
finite-volume power counting. It must be retained whenever the complete
process-dependent contribution enters at an order included in the
finite-volume expansion or is numerically relevant at the target
precision.

Numerical evaluations for $s=0,1$ and $l=0,4$ demonstrate the dependence
on $\nu$ and $\eta$ and distinguish the small-$\nu$ behavior of
the cases containing $E_{n}$ and $1/E_{n}$. The polynomial and
non-polynomial terms can have opposite signs, resulting in a complete
sum$-$integral term whose magnitude is smaller than that of either
component in the examples shown. The numerical results thus illustrate
the mass, momentum, and partial-wave dependence of the dimensionless
functions. Their contribution to a specified observable is obtained
from the linear combination of all relevant propagator structures
after restoring the process-dependent interaction, normalization,
channel, and angular-projection factors. For applications to
finite-volume QCD+QED, the momentum sums must be defined using the
infrared regulator and mode prescription of the selected finite-volume
formulation. Incorporating these functions into complete hadronic
loop calculations will allow the on-shell subtraction term to be
evaluated together with the other finite-volume contributions for
a given process.

\section*{Acknowledgements}
The authors thank Jia-Jun Wu and Akaki Rusetsky for helpful discussions and valuable advice.

\appendix

\section{Partial-Wave Expansion in a Cubic Volume}\label{app:cubic-partial-waves}

At vanishing total momentum, the spatial symmetry group of a periodic cubic
volume is the cubic group $O_{h}$ \cite{ref25,ref26,ref27,ref28,ref35}.
Therefore, the continuous $SO(3)$ angular-momentum multiplet for a fixed $l$
must be organized according to its transformation properties under $O_{h}$.
We express the angular dependence in partial-wave form as

\begin{equation}
\mathcal{Y}_{lm}(\mathbf{n})=n^{l}Y_{lm}(\hat{\mathbf{n}}),\qquad n=|\mathbf{n}|,
\end{equation}
with the same normalization convention as in the main text.

For a sum over complete momentum shells with a radial weight, the magnetic components for fixed $l$ are related by $O_{h}$ symmetry \cite{ref26}. Because inversion is an element of $O_{h}$, all odd-$l$ shell sums vanish. For $l<12$, invariant combinations with even $l$ occur for $l=0,4,6,8,10$, whereas the complete-shell sum vanishes for $l=2$. Thus, $l=4$ is the lowest higher angular component allowed by cubic symmetry and is the higher partial wave considered in the numerical examples of Sec.~4.

For the even values of $l$ listed above, invariance under a quarter turn about a cubic axis restricts the magnetic quantum numbers to $m=0,\pm4,\pm8,\ldots$. The exact relations for the finite-volume zeta functions quoted in Ref. \cite{ref26} follow from this complete-shell symmetry. Because the corresponding infinite-volume integral is rotationally invariant, the sum$-$integral differences defined in Eq.~\eqref{eq:Dg-general} satisfy

\begin{subequations}\label{eq:cubic-magnetic-relations}
\begin{align}
\mathcal{D}_{4,\pm4}(\eta,\nu) & =\sqrt{\frac{5}{14}}\,\mathcal{D}_{4,0}(\eta,\nu),\label{eq:cubic-l4-m4}\\
\mathcal{D}_{6,\pm4}(\eta,\nu) & =-\sqrt{\frac{7}{2}}\,\mathcal{D}_{6,0}(\eta,\nu),\label{eq:cubic-l6-m4}\\
\mathcal{D}_{8,\pm4}(\eta,\nu) & =\sqrt{\frac{14}{99}}\,\mathcal{D}_{8,0}(\eta,\nu),\label{eq:cubic-l8-m4}\\
\mathcal{D}_{8,\pm8}(\eta,\nu) & =\sqrt{\frac{65}{198}}\,\mathcal{D}_{8,0}(\eta,\nu),\label{eq:cubic-l8-m8}\\
\mathcal{D}_{10,\pm4}(\eta,\nu) & =-\sqrt{\frac{66}{65}}\,\mathcal{D}_{10,0}(\eta,\nu),\label{eq:cubic-l10-m4}\\
\mathcal{D}_{10,\pm8}(\eta,\nu) & =-\sqrt{\frac{187}{130}}\,\mathcal{D}_{10,0}(\eta,\nu).\label{eq:cubic-l10-m8}
\end{align}
\end{subequations}

The use of $m=0$ in the numerical evaluation fixes the normalization
of the $O_{h}$-related components. The allowed components with $m\neq0$
are determined by Eqs.~\eqref{eq:cubic-magnetic-relations}, whereas the general expressions retain
the higher partial waves. This reduction requires complete $O_{h}$
shells.

\section*{AI-Assisted Preparation Statement}

Parts of the manuscript were edited using AI-assisted academic writing
tools under the authors' direction. The authors are responsible for all
arguments, derivations, citations, numerical results, and the final wording. 

\end{document}